# Giant Photonic Response of Mexican-hat Topological Semiconductors for Mid-infrared to THz Applications

Haowei Xu[1], Jian Zhou[1], Hua Wang[1] and Ju Li[1,2] [†]

[1] Department of Nuclear Science and Engineering, Massachusetts Institute of Technology, Cambridge, Massachusetts 02139, USA

[2] Department of Materials Science and Engineering, Massachusetts Institute of Technology, Cambridge, Massachusetts 02139, USA

## Abstract

The mid-infrared (MIR), far-infrared (FIR) to terahertz (THz) frequencies are the least developed parts of the electromagnetic spectrum for applications. Traditional semiconductor technologies like laser diodes and photodetectors are successful in the visible light range, but are still confronted with great challenges when extended into the MIR/FIR/THz range. In this paper, we demonstrate that topological insulators (TIs), especially those with Mexican-hat band structure (MHBS), provide a route to overcome these challenges. The optical responses of MHBS TIs can be one to two orders of magnitude larger than that of normal semiconductors at the optical-transition edge. We explore the databases of topological materials and discover a number of MHBS TIs whose bandgaps lie between $0.05 \sim 0.5$ eV and possess giant gains (absorption coefficients) on the order of $10^4 \sim 10^5$ cm$^{-1}$ at the transition edge. These findings may significantly boost potential MIR/FIR/THz applications such as photon sources, detectors, ultrafast electro-optical devices, and quantum information technologies.

[†] correspondence to: liju@mit.edu

The mid-infrared (MIR), far-infrared (FIR) to terahertz (THz) frequencies are of significant importance for applications such as non-destructive inspection, communication, astrophysics, etc. However, lying between the microwave and visible frequencies of the electromagnetic spectrum, the MIR/FIR/THz frequency range is not yet sufficiently developed for many applications, limited by both photon generation and detection technologies. For photon generation, quantum cascade lasers (QCL)[1–3] that rely on the inter-subband transition of electrons in semiconductor quantum wells have become the leading coherent MIR radiation source. However, when extending QCLs to the MIR/FIR/THz range, there are still great challenges due to the physical limitation on the standard techniques of injecting electrons into the high-energy levels of the quantum well. Other mechanisms for MIR/FIR/THz generation are also under active research[4–10]. On the photon detection side[11,12], a variety of physical mechanisms can be utilized, including the thermal effect of MIR/FIR/THz radiations in e.g. a Golay cell, or the interaction between MIR/FIR/THz photons and electrons in e.g. a Schottky barrier diode. These detectors have their pros and cons and are suitable for different applications. In general, detectors with small noise-equivalent power and large modulation frequency are desirable. Single THz photon detectors are also necessary for certain applications[13,14].

In the visible range, the semiconductor laser diodes and photodetectors are widely used. They utilize the same mechanism of continuum-to-continuum (C2C) transition of an electron between the valence band (VB) and conduction band (CB), and the photon generation and detection can be regarded as reciprocal processes, sharing the same pre-factor in the Fermi's Golden rule. However, in the MIR/FIR/THz range, attention on semiconductor devices is mostly focused on the transition between bound-to-continuum (B2C) or bound-to-bound (B2B) states, in e.g. an extrinsically doped semiconductor or a quantum well. Such B2C and B2B transitions impose stringent requirements on material fabrications, such as precise control on the impurity type and concentration. C2C transition actually has huge advantages over B2C or B2B. For example, the carrier mobilities would not be curtailed by dopants, and the detector may be utilized under arbitrary polarizations. However, 1 THz is equivalent to 4.1 meV, far below the typical interband gaps of most known semiconductors (on the order of 1 eV). Another significant bottleneck is that the optical transition rate reduces drastically with decreasing photon frequency. In free space the dipole radiation power scales as $\omega^4$, where $\omega$ is the angular frequency of the photon. Even if an optical cavity is used to modify the photon density of states, the radiation power still scales at least with $\omega$, thus for small $\omega$ the photon emission or absorption would be slow. A third obstacle is that when the bandgap of the semiconductor is below the frequency of optical phonons (usually on the order of 1-10 THz), the electron-phonon coupling will induce rapid non-radiative interband transition, which is a dominant source of noise and loss, especially at high temperature. Of course, the non-radiative transition via phonon can be kept slow as long as the bandgap is above the optical phonon frequencies. In fact, the low optical transition rate and non-radiative transition via phonon coupling are two generic problems in all MIR/FIR/THz technologies and are not specific to the C2C transition.

In this work, we propose that topological insulators[15] (TIs)[16–18] could provide opportunities for resolving these issues. The bandgaps of many TIs fall within the MIR/FIR/THz range. And as we will show in the following, the optical responses of TIs can be significantly strongerthan that of normal insulators (NIs) due to enhanced Berry connections. Provided with Mexican-hat band structure (MHBS), the optical responses would be further boosted due to the larger joint density of states (jDOS). Consequently, the optical responses of MHBS TIs in the MIR/FIR/THz range can be one or two orders of magnitude larger than that of the widely used MCT (HgCdTe) alloy. During the past decade, TIs have been under intense research and have changed our understandings of the state of matter. Many novel properties of TIs such as symmetry protected edge states could find potential applications in electronic and spintronic devices[19,20], fault-tolerant quantum computers[21,22], etc. From a band-structure point of view, the non-trivial band topology of TIs usually comes from the inversion of energy bands with respect to their natural energy order[18,23–25]. For example, the metal $d$ orbitals usually constitute higher energy bands than the chalcogen $p$ orbitals. However, in some materials, the metal $d$ band could move below the chalcogen $p$ band for some range of the wavevector $k$. In this case the metal $d$ band would become valence band rather than conduction band[25], and would cause band crossings. A bandgap is opened at the band crossings under spin-orbit coupling (SOC), which alters the original space-group symmetry to double group in spin-wavefunction space. The SOC induced bandgap in TIs brings several advantages: First, the bandgap is usually direct, which could enable fast and efficient optical transitions and eliminate inefficient phonon coupling required for indirect bandgap transitions. Second, the bandgap $E_\text{g}$ is determined by the strength of SOC, and is usually on the order of 0.1 eV or below, lying in the range of MIR/FIR/THz. Furthermore, the bandgap can be controllably and continuously tuned by external field parameters ($\{\lambda_i\}$) such as strain[26–28], composition[29–31], or electric field[25,32,33]. Particularly, varying $\{\lambda_i\}$ can trigger topological phase transitions between topologically non-trivial and trivial phases. At some critical points $\{\lambda_i^c\}$, the system would be gapless[34–37]. Consequently, arbitrarily small bandgaps can be realized around $\{\lambda_i^c\}$. In fact, the alloy widely used for infrared detection, $Hg_xCd_{1-x}Te$, takes advantage of the band inversion as well[38,39]. The system has an inverted bandgap when $x = 1$, and a normal bandgap when $x = 0$. In between, a zero bandgap is achievable by tuning the composition $x$.

Striking advantages of using TIs for MIR/FIR/THz applications can be revealed when considering the bulk optical transitions in TIs. In fact, the optical response of TIs at the transition edge ($\hbar\omega = E_g$) can be one or two orders of magnitude larger than that of NIs. In order to illustrate this point, here we study an important parameter for optical applications, the gain $G(\omega)$ in the laser mode, or equivalently, the absorption coefficient $\mu(\omega)$ in the detector mode. $G(\omega)$ can be related to the dielectric response function $\epsilon(\omega)$ of the semiconductor by $\mu(\omega) = G(\omega) = \frac{2\omega \, \text{Im}(\sqrt{\epsilon(\omega)})}{c}$. The dielectric response function within the random phase approximation (RPA)[40] can be expressed as

$$\epsilon_{ij}(\omega) = \delta_{ij} - \frac{e^2}{\epsilon_0} \int \frac{d^3\mathbf{k}}{(2\pi)^3} \sum_{c,v} \frac{\langle u_{v,\mathbf{k}}|\nabla_{k_i}|u_{c,\mathbf{k}}\rangle \langle u_{c,\mathbf{k}}|\nabla_{k_j}|u_{v,\mathbf{k}}\rangle}{E_{c,\mathbf{k}} - E_{v,\mathbf{k}} - \hbar\omega - i\xi}, \tag{1}$$

where $|u_{n,\mathbf{k}}\rangle$ and $E_{n,\mathbf{k}}$ are the periodic part of the wavefunction and the energy eigenvalue at band $n$ and wavevector $\mathbf{k}$. $c$ and $v$ denote CB and VB, respectively. $\delta_{ij}$ is the Kronecker delta, $i, j = x, y, z$ are Cartesian indices, and $\epsilon_0$ is the vacuum permittivity. Close to the transition edge, only the lowest CB and highest VB need to be considered. From Eq. (1) one can see that two factors have significant influence on $\epsilon(\omega)$. First, the product $\mathcal{O}^{ij}(\mathbf{k}) \equiv |A_{cv}^i(\mathbf{k})A_{vc}^j(\mathbf{k})|$, where $\mathbf{A}_{mn}(\mathbf{k}) \equiv i\langle u_{m,\mathbf{k}}|\nabla_\mathbf{k}|u_{n,\mathbf{k}}\rangle$ is the Berry connection. In order to have large Berry connection, the wavefunctions of CB and VB should have large mixing. The second factor is the joint density of states (jDOS) $\rho_{cv}(\omega)$, which comes into play via the Brillouin zone (BZ) integration. In principle, both $\mathcal{O}(\mathbf{k})$ and $\rho_{cv}(\omega)$ are determined by the band characteristics, including the energy dispersion and the orbital contributions in each band. The details of the band characteristics are material dependent, but the essence of band inversion, a telltale feature of TIs, can be understood with the help of a two-band model Hamiltonian[23,24,41]. The basis wavefunctions are denoted as $|\psi_1\rangle$ and $|\psi_2\rangle$, which are usually orthogonal and have small mixing. The Hamiltonian can be written as

$$H(\boldsymbol{k}) = \begin{bmatrix} M - B_+|k_+|^2 - B_z k_z^2 & A_+ k_+ + A_z k_z \\ A_+ k_+^* + A_z k_z & -(M - B_+|k_+|^2 - B_z k_z^2) \end{bmatrix}. \tag{2}$$

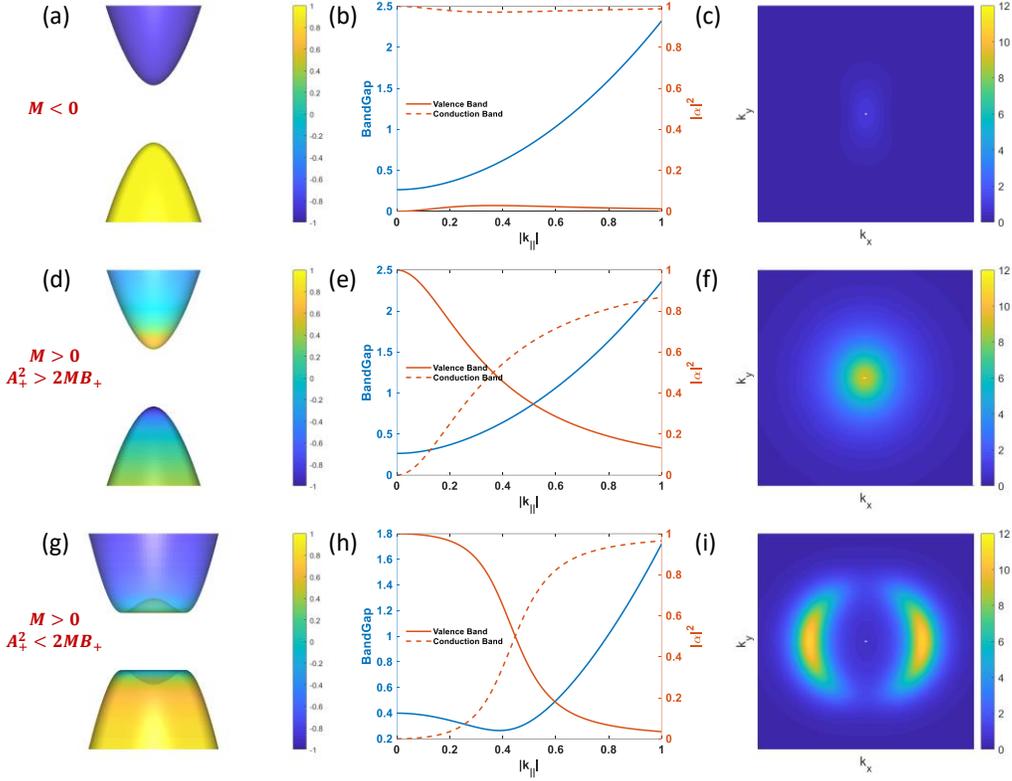

**Figure 1** The band characteristics of the two-band model in Eq. (2). The first column shows the band dispersion in the $k_x$-$k_y$ plane. The color map shows $|\alpha|^2 - |\beta|^2$, which is the difference between the contributions from two basis wavefunctions. The second column shows the bandgap and $|\alpha_{VB,CB}|^2$ along the radial direction of the $k_x$-$k_y$ plane. The third row shows the interband mixing $\mathcal{O}^{xx}(\boldsymbol{k})$ on the $xy$ plane. The first row corresponds to a normal insulator without band inverison, while the second and third row have band inverisons. Particularly, the third row exhibits MHBS.

Here we assume some basic crystal symmetries (e.g. in-plane three-fold or four-fold rotational symmetry) so that the mixing terms for in-plane wavevectors $k_{x,y}$ are identical and are denoted by $A_+$ and $B_+$. $k_+ \equiv k_x + ik_y$ is the in-plane complex momentum. $M$ characterizes the band inversion. By solving Eq. (2), one obtains the band dispersion

$$E_\pm(\boldsymbol{k}) = \pm\sqrt{(M - B_+|k_+|^2 - B_z k_z^2)^2 + |A_+ k_+ + A_z k_z|^2} \tag{3}$$

The wavefunctions are $|\text{VB}\rangle = \alpha_{\text{VB}}(\boldsymbol{k})|\psi_1\rangle + \beta_{\text{VB}}(\boldsymbol{k})|\psi_2\rangle$, $|\text{CB}\rangle = \alpha_{\text{CB}}(\boldsymbol{k})|\psi_1\rangle + \beta_{\text{CB}}(\boldsymbol{k})|\psi_2\rangle$.

Without loss of generality, we set $k_z = 0$ and focus on the $k_x$-$k_y$ plane. Some band characteristics on the $k_x$-$k_y$ plane are plotted in Figure 1. The three columns are $E(\mathbf{k})$, $|\alpha_{\text{VB,CB}}(\mathbf{k})|^2$ and $\mathcal{O}^{xx}(\mathbf{k})$, respectively. The panels a-c corresponds to $M < 0$, while the second (panels d-f) and the third (panels g-i) rows have positive $M$. For all three rows the parameters are adjusted so that the direct bandgaps are identical. When $M$ is negative, the mass term $M - B_+|k_+|^2 < 0$, and the VB and CB are dominantly composed of $|\psi_2\rangle$ and $|\psi_1\rangle$ for all $k_+$ (Figure 1b). There is no band inversion, corresponding to a topologically trivial state. $\mathcal{O}^{xx}(\mathbf{k})$ in this case is plotted in Figure 1c, where we can see that it is quite small. On the other hand, when $M$ is positive, band inversion occurs, and the band dispersion has different shapes depending on the magnitude of $M, A_+$ and $B_+$ (Figure d and g). When $A_+^2 > 2MB_+$ (Figure 1d-e), the band dispersion is nearly parabolic, with the band edge lying at $\mathbf{k} = 0$. While for $A_+^2 < 2MB_+$ (Figure 1f-i), the band dispersion shows a MHBS, with the band edge forming a ring at finite $|k_+|$. The wavefunction compositions undergo an inversion for both cases: At a small $|k_+|$, the mass term is positive, and the VB and CB are dominated by $|\psi_1\rangle$ and $|\psi_2\rangle$, respectively. On the other hand, when $|k_+|$ is large, the mass term would turn negative, and the orbital contributions to the two bands are inverted, with the VB and CB dominated by $|\psi_2\rangle$ and $|\psi_1\rangle$, respectively. Such a band inversion usually indicates a non-trivial band topology. In between these two extremes, VB and CB have comparable contributions from both $|\psi_1\rangle$ and $|\psi_2\rangle$. Owing to the band inversion, $\nabla_{\mathbf{k}}\alpha(\mathbf{k})$ and $\nabla_{\mathbf{k}}\beta(\mathbf{k})$ are larger than those in the topologically trivial case, and the wavefunction overlap is also stronger. Consequently, $\mathcal{O}(\mathbf{k})$ is supposed to be larger, which is verified by the numerical results in Figure 1f and 1i, where we can see that the maximum value of $\mathcal{O}(\mathbf{k})$ and its $\mathbf{k}$-space distribution are much larger than that in Figure 1c. Note that $\mathcal{O}(\mathbf{k})$ does not appear to be isotropic in Figures 1c,f,i because only $\mathcal{O}^{xx}(\mathbf{k})$ is plotted; the isotropy would be restored when $\mathcal{O}^{yy}(\mathbf{k})$ contribution is added.

As stated above, the band inversion alone can contribute to stronger and faster interband transition. This would already lead to stronger optical responses for both parabolic and MHBS TIs (see Supporting Information, SI). However, the MHBS brings in an additional and even more striking advantage: greater jDOS at the transition edge. The jDOS can be written as

$$\rho_{cv}(\omega) = \frac{2}{(2\pi)^d} \int_{E_c - E_v = \hbar\omega} \frac{dS}{|\nabla_k (E_c - E_v)|} \tag{4}$$

where $d$ is the dimensionality of the system and the integration is over the constant energy-difference surface (CEDS) with $E_c - E_v = \hbar\omega$ in the first BZ. For a standard parabolic band structure, it is well-known that in 3D, $\rho_{cv}(\omega)$ scales as $\sqrt{\hbar\omega - E_g}$ and is zero at the transition edge. Whereas in 2D, $\rho_{cv}(\omega)$ should be a constant. However, with a MHBS as in Eq. (3), it can be shown that at transition edge, $\rho_{cv}(E_g)$ has a finite value in 3D and is infinite in 2D. This can be understood by a geometric analysis: for a parabolic band structure, the CEDS is a 0D point at the transition edge, whereas for a MHBS, the CEDS forms a 1D closed loop, which boosts $\rho_{cv}(E_g)$ from zero to a finite value in 3D, and from a finite value to infinity in 2D. Consequently, in 3D the gain $G(\omega)$ would abruptly jump to a finite value, rather than starting from zero and increasing smoothly at the transition edge. While in 2D, the gain would exhibit a sharp peak at the transition edge (see SI).

We take 3D cubic SnSe (space group $Fm\overline{3}m$) as an example to illustrate our theory. The atomic structure is shown in the inset of Figure 2b. Previous studies have demonstrated that SnSe is a topological crystalline insulator (TCI)[42,43] with fully relaxed lattice constant $a = b = c = 6.07$ Å. Here the band structure, band topology and optical properties are calculated with density functional theory (DFT) with Perdew-Burke-Ernzerhof (PBE)[44] functional. Results with Heyd, Scuseria, and Ernzerhof (HSE)[45,46] hybrid functional can be found in the SI. The band structure is shown in Figure 2a, where the fundamental bandgap at L-point is inverted with a magnitude of $-0.53$ eV. Then the SOC opens an inverted bandgap of $E_g = 0.21$ eV around L point. The first BZ of cubic SnSe is a truncated octahedron (inset of Figure 2c) and L point is the center of the hexagonal (111) faces (painted in cyan). On the hexagonal face, the system has a three-fold rotational symmetry. Therefore, the Hamiltonian Eq. (2) can be applied here: $x$-$y$ plane is on the hexagonal surface, while $z$ is along the $\Gamma - L$ direction. As expected, the inverted bandgap forms a ring on the hexagonal face, and the jDOS abruptly jumps to a finite value at the transition edge (Figure 2b). The imaginary part of the dielectric function $\epsilon_{xx}^{(2)}(\omega)$ is plotted in Figure 2c. There is a strong peak around $E_g$, which is rarely seen in the dielectric response of normal 3D insulators. As discussed above, the peak should be attributed to 1) Strong VB and CB wavefunction mixing due to band inversion, and 2) Large jDOS around $E_g$ owing to the MHBS. Finally, the gain $G(\omega)$ (Figure 1d) jumps to a tremendous value of $4 \times 10^4$ cm$^{-1}$ at $E_g \sim 0.2$ eV, which is comparable magnitude-wise to that of GaAs in the visible-light range[47].

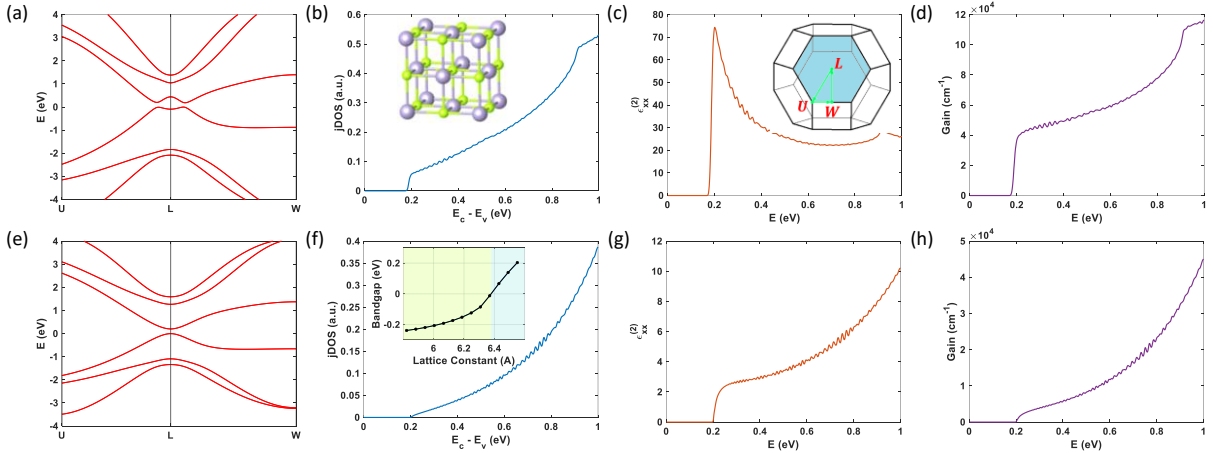

**Figure 2** The band structure and optical properties of cubic SnSe. The four columns correspond to band structure, jDOS, imaginary part of the dielectric function $\epsilon_{xx}^{(2)}$, and the optical gain $G(\omega) = 2\omega\, Im(\sqrt{\epsilon(\omega)})/c$, respectively. For the first row, the atomic structure of SnSe is fully relaxed, and it is a topological crystalline insulator (TCI) with MHBS. The second row corresponds to the SnSe under an 8% triaxial strain, which is normal insulating without band inversion. The dramatic improvement over optical properties by the band inversions and MHBS can be seen by comparing two rows. Inset of (b): atomic structure of cubic SnSe. Inset of (c): first BZ of cubic SnSe. The bandedge is on the L point and the MHBS is on the hexagonal surface in cyan color. Inset of (f): bandgap of cubic SnSe as a function of lattice constant $a$. With $a \lesssim 6.4$ Å, SnSe is a TCI with inverted bandgap. While for $a \gtrsim 6.4$ Å, it is a normal insulator. A negative bandgap indicates that it is an inverted bandgap.

In order to fully compare the salient merits of the MHBS topological materials with that of NIs, we elastically strained the lattice parameter of SnSe while keeping its cubic phase, so that it undergoes a topological phase transition and becomes a NI. The bandgap variation as a function of strain is shown in the inset of Figure 2f. The phase transition happens around $a = 6.4$ Å. When the lattice constant $a \lesssim 6.4$ Å, SnSe remains topologically nontrivial, whereas for $a \gtrsim 6.4$ Å, the inverted bandgap disappears and SnSe becomes a NI. Furthermore, when $a = 6.55$ Å (corresponding to a 8% tensile strain), the normal bandgap is also 0.2 eV, close to the inverted bandgap in the strain-free state. Thus, we calculate the optical properties of SnSe with $a = 6.55$ Å. The results (Figures 2e-h) exhibit typical features of a NI: The jDOS, dielectric response and gain all have slow risings from zero at $E_g$. Even at 0.3 eV, the gain is still only $4 \times 10^3$ cm$^{-1}$, ten times smaller than that in the strain-free topological case at $E_g = 0.2$ eV (Figure 2d). This demonstrates

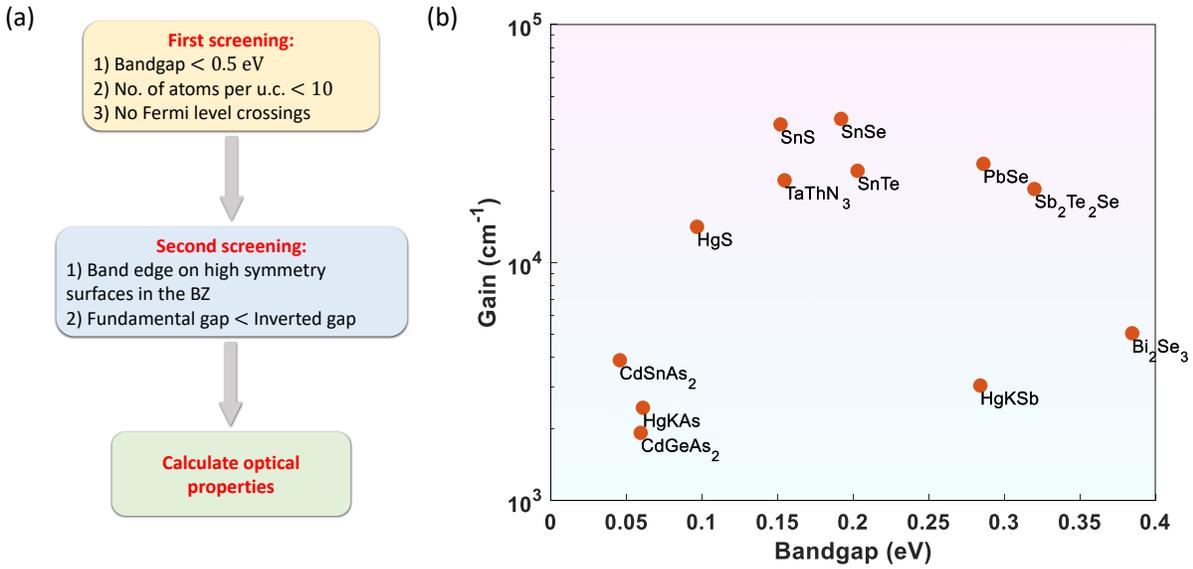

**Figure 3** (a) workflow chart of the exploration of materials with enhanced optical properties in the MIR/THz range in the topological material database. (b) Scattering plot of bandgap and gain of materials targeted by the workflow in (a).

that optical response and its potential applications (such as laser mode gain and detector absorption) can be greatly enhanced in topological materials (TI or TCI) with MHBS by one or two orders of magnitude. Therefore, thin-film photonic devices made with such Mexican-hat TIs can be much thinner than those made with normal semiconductors, while maintaining high photon detection efficiency and/or emissivity.

Finally, in order to search for more topological materials for MIR/FIR/THz applications, we explored the entire databases of topological material[48,49] and calculated the optical properties of TIs with MHBS. One should note that the database is constructed with PBE functional, and would be different if HSE functional is used. Here we focus on the PBE database for the present illustrative work. We first mapped out TIs with bandgap lower than $0.5$ eV ($\approx 120$ THz), number of atoms per unit cell fewer than 10, and without any Fermi level crossings. Then we carried out a screening by investigating the crystalline symmetry and band structures of these TIs, and targeted candidates that could potentially have MHBS over at least one surface in the BZ. Finally, we calculated the optical properties of these candidate TIs. The results are plotted in Figure 3b (see also Table 1), where we can see a number of TIs with $G(\omega)$ over $10^4$ cm$^{-1}$ at transition edge ranging between $0.1 \sim 0.5$ eV. There are also three TIs HgKAs, CdSnAs$_2$, CdGeAs$_2$, whose bandgaps are around $0.05$ eV and $G(E_g)$ are on the order of several $10^3$ cm$^{-1}$. In order to eliminate possible photon-phonon transition, materials with too small bandgaps (below 10 THz) are not considered here. Note that because the optical transition rate is slower for small $\omega$, it is harder to obtain large $G(\omega)$ at small $\omega$ (also manifested by the relation $G(\omega) = \frac{2\omega \text{Im}(\sqrt{\epsilon(\omega)})}{c}$). These TIs should be compared with the widely used infrared detector, $\text{Hg}_x\text{Cd}_{1-x}\text{Te}$ (MCT) alloy. The gain/absorption coefficient of $\text{Hg}_x\text{Cd}_{1-x}\text{Te}$ is only around $10^3$ cm$^{-1}$ near the transition edge[50]. In other words, the performance of the TIs as listed in Figure 3b is at least ten times better than that of $\text{Hg}_x\text{Cd}_{1-x}\text{Te}$ in terms of gain/absorption coefficient, and should be of primary technological interest.

Before concluding, we would like to make several remarks. First, the enhancement of optical response in Mexican hat TIs is a bulk effect, rather than surface effects that are widely studied for TIs. Thus, the enhancement of optical responses in TIs with MHBS are applicable to 2D materials as well, which has bulk interaction with light. As discussed before, in addition to the large interband wavefunction mixing, the jDOS of 2D Mexican hat materials can be infinity at the transition edge. As a result, the absorbance of 2D Mexican hat TIs can be dramatically improved, which is verified with 2D monolayer SnF[51] as shown in the SI. Furthermore, quantum wells constructed with 3D materials are effectively 2D materials due to the confinement in the $z$ direction, thus they should exhibit similar properties as well, which is advantageous for applications like laser diodes.

Second, the 2D surfaces of 3D TIs are semi-metallic with vanishing bandgap. Therefore, they could affect the performance of TIs as light sources or detectors by providing a route for non-radiative recombination of electrons and holes in the laser mode, or reflecting the incoming light in the detector mode. However, these effects should be relatively weak compared to the bulk effect in the frequency range of interest for the following reasons. First, the surface states are actually *semi*-metallic, meaning that the free carrier concentration on the surface is low. In practice, there could be free carriers from e.g., doping or thermal activation. But the density of these free carriers $n$ should be smaller by several orders of magnitude than that in real metals such as copper, which is on the order of $10^{28}$ m$^{-3}$. Assuming the surface free carrier concentration is $n = 10^{23}$ m$^{-3}$ within the penetration depth (~1 nm) of the surface states, the free carrier contribution to the dielectric function of SnSe is only on the order of 100 even at $\omega = 1$ THz from Drude model (see SI), which is smaller than that of copper by 5 orders of magnitude. Since the interband contribution is also ~100, the total dielectric function of the surface states is on the order of 100 within the ~1 nm skin depth, which is not so big. Therefore, the absorption and reflection from surface states are much less efficient than that in real metals. Also, with $n = 10^{23}$ m$^{-3}$, the surface carriers plasmon frequency is about 2 THz ~ 8 meV, well below the bandgap of the bulk material, which is the frequency region of interest. Second, the surface states only exist near the surface with a small (joint) DOS. The thickness of these surface states is on the order of 1 nm, which is much smaller than the wavelength of MIR/FIR/THz photons and the size of typical bulk materials. In the detector mode, the surface states might reflect the incoming light. However, to obtain full reflection, the penetration depth of light upon reflection needs to be comparable with its wavelength, which is on the order of 0.1 ~ 1 μm with $\epsilon \approx 100$. This is much larger than the thickness of the surface states. Therefore, these surfaces states are too thin to reflect the incoming light strongly, and thus light can penetrate into the bulk. For example, monolayer graphene, despite being a Dirac metal, allows 97.4% light transmission. Actually, experimental works have shown that the total reflectivity of several TIs is about 0.5 near the bulk bandgap[52–54]. Considering that the total reflectivity includes the contribution from the bulk, the reflectivity contribution from surface electronic states should be less significant. In the laser mode, the bulk states are extended over the entire bulk. By a simple size analysis, one can see that the (joint) DOS of the bulk states is much larger than that of the surface states, whose thickness is only about 1 nm. Therefore, the non-radiative recombination on the surfaces should be weak as compared with the radiative recombination in the bulk due to the small (joint) DOS. Furthermore, the stimulated amplification would drastically accelerate the radiative recombination in the bulk, rendering the influence of the surface states even weaker. More detailed discussions on the influence of the surface states can be found in the SI.

Third, the optical response of TIs in Figure 3b are boosted by the combination of two factors, namely, the large Berry connection and the large jDOS. But, each factor alone could contribute to an enhanced optical response. Therefore, a common TI with parabolic band structure could also have augmented optical response over NIs. In fact, with the model Hamiltonian Eq. (2), the imaginary part of the dielectric function $\epsilon^{(2)}$ of the TI without MHBS is about 3 times larger than that of the NI near the transition edge (see SI), due to TI induced band mixing (compare Fig. 1e to Fig. 1b). In addition, we only explored TIs in the databases of topological materials, but the band inversion does not necessarily lead to nontrivial band topology, thus materials outside the topological materials database could also have inverted band structure and Mexican-hat features. Thus, many other materials with MHBS and/or band-mixing features could have enhanced optical response that could enable efficient MIR/FIR/THz applications. The discovery and characterization of those materials shall be discussed elsewhere.

Finally, in this work we only studied the linear optical responses. Actually, the nonlinear optical responses, such as the second harmonic generation and bulk photovoltaic effect, could also be boosted in TIs, especially those with MHBS.

In conclusion, we have demonstrated giant continuum-to-continuum (C2C) optical responses at the conduction-to-valence band transition edge of topological materials with MHBS. By investigating their wavefunction characteristics, we attribute the giant optical response to two factors, namely the larger jDOS, and stronger wavefunction mixing due to band inversion. We explored the entire databases of topological materials and discovered a number of TIs with remarkable photonic responses, which could enable efficient MIR/FIR/THz photon generation, detection and manipulation.

**Table 1** Materials in Figure 3d. The results are from DFT calculations with PBE functional. The band topology of these materials with HSE hybrid functional can be found in the SM

| Material | ICSD | Space group | Lattice constant (Å) | Bandgap $E_g$ (eV) | Gain ($\times 10^3$ cm$^{-1}$) |
|---|---|---|---|---|---|
| $Bi_2Se_3$ | 617072 | 166 ($R\bar{3}m$) | $a = b = 4.39, c = 30.50$ | 0.385 | 5.04 |
| $Sb_2Te_2Se$ | 2085 | 166 ($R\bar{3}m$) | $a = b = 4.19, c = 29.94$ | 0.320 | 20.4 |
| $CdGeAs_2$ | 16736 | 122 ($I\bar{4}2d$) | $a = b = 5.94, c = 11.22$ | 0.059 | 1.92 |
| $CdSnAs_2$ | 16737 | 122 ($I\bar{4}2d$) | $a = b = 6.09, c = 11.92$ | 0.046 | 3.88 |
| HgKSb | 56201 | 194 ($P6_3/mmc$) | $a = b = 4.78, c = 10.23$ | 0.284 | 3.04 |
| HgKAs | 10458 | 194 ($P6_3/mmc$) | $a = b = 4.51, c = 9.98$ | 0.061 | 2.45 |
| HgS | 56476 | 216 ($F\bar{4}3m$) | $a = b = c = 5.85$ | 0.097 | 14.1 |
| PbSe | 238502 | 225 ($Fm\bar{3}m$) | $a = b = c = 6.01$ | 0.286 | 26.0 |
| SnS | 651015 | 225 ($Fm\bar{3}m$) | $a = b = c = 5.80$ | 0.152 | 38.0 |
| SnSe | 52424 | 225 ($Fm\bar{3}m$) | $a = b = c = 5.99$ | 0.192 | 40.1 |
| SnTe | 52489 | 225 ($Fm\bar{3}m$) | $a = b = c = 6.32$ | 0.203 | 24.3 |
| $TaThN_3$ | 77661 | 221 ($Pm\bar{3}m$) | $a = b = c = 4.02$ | 0.155 | 22.2 |

**Methods** The first-principles calculations are based on density functional theory (DFT)[55,56] as implemented in Vienna *ab initio* simulation package (VASP)[57,58]. Generalized gradient approximation (GGA) of Perdew-Burke-Ernzerhof (PBE)[44] form is used to treat the exchange-correlation interactions. Core and valence electrons are treated by projector augmented wave (PAW) method[59] and a plane wave basis set, respectively. For the DFT calculations, the first Brillouin zone is sampled by a Γ-centered ***k*-**mesh with grid density of at least $2\pi \times 0.02$ Å$^{-1}$ along each dimension. The Bloch waves from DFT calculations are projected onto the maximally-localized Wannier functions (MLWF) with the Wannier90 package[60]. Then a tight-binding (TB) Hamiltonian is constructed from the MLWF. And the TB Hamiltonian is used to calculate the dielectric function in Eq. (1) on a much denser ***k*** -mesh with grid density of at least $2\pi \times 0.001$ Å$^{-1}$ along each dimension.

## Acknowledgements

This work was supported by an Office of Naval Research MURI through grant #N00014-17-1-2661.

(43) Hsieh, T. H.; Lin, H.; Liu, J.; Duan, W.; Bansil, A.; Fu, L. Topological Crystalline Insulators in the SnTe Material Class. *Nat. Commun.* **2012**, *3* (1), 1–7.

(44) Perdew, J. P.; Burke, K.; Ernzerhof, M. Generalized Gradient Approximation Made Simple. *Phys. Rev. Lett.* **1996**, *77* (18), 3865–3868.

(45) Heyd, J.; Scuseria, G. E.; Ernzerhof, M. Hybrid Functionals Based on a Screened Coulomb Potential. *J. Chem. Phys.* **2003**, *118* (18), 8207–8215.

(46) Heyd, J.; Scuseria, G. E.; Ernzerhof, M. Erratum: "Hybrid Functionals Based on a Screened Coulomb Potential" [J. Chem. Phys. 118, 8207 (2003)]. *J. Chem. Phys.* **2006**, *124* (21), 219906.

(47) Casey, H. C.; Sell, D. D.; Wecht, K. W. Concentration Dependence of the Absorption Coefficient for N- and p-Type GaAs between 1.3 and 1.6 EV. *J. Appl. Phys.* **1975**, *46* (1), 250–257.

(48) Bradlyn, B.; Elcoro, L.; Cano, J.; Vergniory, M. G.; Wang, Z.; Felser, C.; Aroyo, M. I.; Bernevig, B. A. Topological Quantum Chemistry. *Nature* **2017**, *547* (7663), 298–305.

(49) Vergniory, M. G.; Elcoro, L.; Felser, C.; Regnault, N.; Bernevig, B. A.; Wang, Z. A Complete Catalogue of High-Quality Topological Materials. *Nature* **2019**, *566* (7745), 480–485.

(50) Scott, M. W. Energy Gap in Hg1-XCdxTe by Optical Absorption. *J. Appl. Phys.* **1969**, *40* (10), 4077–4081.

(51) Xu, Y.; Yan, B.; Zhang, H. J.; Wang, J.; Xu, G.; Tang, P.; Duan, W.; Zhang, S. C. Large-Gap Quantum Spin Hall Insulators in Tin Films. *Phys. Rev. Lett.* **2013**, *111* (13), 136804.

(52) Sushkov, A. B.; Jenkins, G. S.; Schmadel, D. C.; Butch, N. P.; Paglione, J.; Drew, H. D. Far-Infrared Cyclotron Resonance and Faraday Effect in Bi 2Se3. *Phys. Rev. B - Condens. Matter Mater. Phys.* **2010**, *82* (12), 125110.

(53) Hada, M.; Norimatsu, K.; Tanaka, S.; Keskin, S.; Tsuruta, T.; Igarashi, K.; Ishikawa, T.; Kayanuma, Y.; Miller, R. J. D.; Onda, K.; et al. Bandgap Modulation in Photoexcited Topological Insulator Bi2Te3 via Atomic Displacements. *J. Chem. Phys.* **2016**, *145* (2), 024504.

(54) Wang, Y.; Luo, G.; Liu, J.; Sankar, R.; Wang, N. L.; Chou, F.; Fu, L.; Li, Z. Observation of Ultrahigh Mobility Surface States in a Topological Crystalline Insulator by Infrared Spectroscopy. *Nat. Commun.* **2017**, *8* (1), 1–8.

(55) Hohenberg, P.; Kohn, W. Inhomogeneous Electron Gas. *Phys. Rev.* **1964**, *136* (3B), B864–B871.

(56) Kohn, W.; Sham, L. J. Self-Consistent Equations Including Exchange and Correlation Effects. *Phys. Rev.* **1965**, *140* (4A), A1133–A1138.

(57) Kresse, G.; Furthmüller, J. Efficiency of Ab-Initio Total Energy Calculations for Metals and Semiconductors Using a Plane-Wave Basis Set. *Comput. Mater. Sci.* **1996**, *6* (1), 15–50.

(58) Kresse, G.; Furthmüller, J. Efficient Iterative Schemes for *Ab Initio* Total-Energy Calculations Using a Plane-Wave Basis Set. *Phys. Rev. B* **1996**, *54* (16), 11169–11186.

(59) Blöchl, P. E. Projector Augmented-Wave Method. *Phys. Rev. B* **1994**, *50* (24), 17953–17979.

(60) Mostofi, A. A.; Yates, J. R.; Pizzi, G.; Lee, Y. S.; Souza, I.; Vanderbilt, D.; Marzari, N. An Updated Version of Wannier90: A Tool for Obtaining Maximally-Localised Wannier Functions. *Comput. Phys. Commun.* **2014**, *185* (8), 2309–2310.
16